\documentstyle[lprocl,11pt]{article}

\def\Journal#1#2#3#4{{#1} {\bf #2}, #3 (#4)}

\def\NPB{{\em Nucl. Phys.} B}
\def\PLB{{\em Phys. Lett.}  B}
\def\PRL{\em Phys. Rev. Lett.}
\def\PRD{{\em Phys. Rev.} D}

\def\be{\begin{equation}}
\def\ee{\end{equation}}
\def\bea{\begin{eqnarray}}
\def\eea{\end{eqnarray}}
\def\<{\langle}
\def\>{\rangle}

\begin{document}

\title{GLUON FRAGMENTATION TO ${}^3D_J$ QUARKONIA AND TEST OF COLOR-OCTET
PRODUCTION MECHANISM}

\author{ KUANG-TA CHAO }

\address{Department of Physics, Peking University,
Beijing 100871, People's Republic of China}

\author{CONG-FENG QIAO}
\address{CCAST (World Laboratory), Beijing 100080, People's Republic of China}

\author{FENG YUAN}
\address{Department of Physics, Peking University,
Beijing 100871, People's Republic of China}

\maketitle\abstracts{
We report the recent progress in the studies of $D$-wave heavy
quarkonia production from gluon fragmentation at the Tevatron and in $Z^0$
decays, and the $D$-wave charmonia production in $B$ decays.
We show that the color-octet contributions are of crucial importance to
the production of $D$-wave quarkonia.
Signals, as strong as that for $J/\psi$ and $\psi^\prime$, should be seen
for the $D$-wave charmonium in the processes.
These will provide a test of the color-octet production
mechanism.}

\section{Introduction}

Studies of heavy quarkonium production in high energy collisions provide important
information on both perturbative and nonperturbative QCD.
In recent years, a new framework for treating quarkonium production 
and decays has been advocated by Bodwin, Braaten and Lepage in the context of 
nonrelativistic quantum chromodynamics (NRQCD) \cite{bbl}. 
In this approach, the production process is factorized into short and long 
distance parts, while the latter is associated with the nonperturbative
matrix elements of four-fermion operators. This factorization formalism
gives rise to a new production mechanism called the color-octet mechanism, in which
the heavy-quark and antiquark pair is produced at short distance in a
color-octet configuration and subsequently
evolves nonperturbatively into physical quarkonium state.
The color-octet term in the gluon fragmentation to $J/\psi$($\psi'$) has been 
considered to explain the $J/\psi$($\psi'$) surplus problems discovered
by CDF \cite{surplus,s1}.
Taking the nonperturbative $\< {\cal O}^{J/\psi}_8(^3S_1)\> $ and
$\< {\cal O}^{\psi'}_8(^3S_1)\> $ as input parameters, the CDF surplus problem
for $J/\psi$ and $\psi'$ can be explained as the contributions of color-octet
terms due to gluon fragmentation.
In the past few years, applications of the NRQCD factorization formalism
to $J/\psi$($\psi^\prime$) production at various experimental facilities
have been studied \cite{annrev}.

Even though the color-octet mechanism has gained some successes in describing 
the production and decays of heavy quark bound systems ,
it still needs more effort to go before finally setting its position and role in
heavy quarkonium physics.
(For instance, the photoproduction data from HERA \cite{photo} put a question
about the universality of the color-octet matrix elements \cite{photo2}, in which
the fitted values of the matrix elements $\langle {\cal O}^{J/\psi}_8({}^1S_0)\rangle$
and $\langle {\cal O}^{J/\psi}_8({}^3P_J)\rangle$ are one order of
magnitude smaller than those determined from the Tevatron data \cite{s1}).

In this report, we discuss the production of $D$-wave heavy quarkonia
under the framework of NRQCD factorization formalism,
including gluon fragmentation at the hadron collider \cite{gf},
in the $Z^0$ decays \cite{z0}, and the $D$-wave charmonia production in $B$ decays
 \cite{bdw}.
All these results show that the color-octet mechanism is crucially 
important to $D$-wave charmonium production,
and the color-octet contributions are found to be over 
two orders of magnitude larger than the color-singlet contributions.
This because, in essence, the above processes are associated with the color-octet
${}^3S_1$ intermediate configuration, which is the same as that in
$J/\psi$ or $\psi^\prime$ production at the Tevatron via gluon fragmentation.
Therefore, the production of $D$-wave quarkonia would provide a crucial test
of the color-octet production mechanism.
On the other hand, even if the color-singlet model predicts the $D$-wave production rates
too small to be visible, they could now be detected after and only after
including the color-octet production mechanism.
It is the color-octet mechanism that could make it possible to search for the $D$-wave
heavy quarkonium states, and then complete the studies of
charmonium and bottomonium families.

\section{Gluon Fragmentation to ${}^3D_J$ Heavy Quarkonia at the Tevatron}

We choose a special process to study the gluon fragmentation to color-singlet
${}^3D_J$ quarkonium,
\be
Q^*\rightarrow Q g^*;~~ g^*\rightarrow {}^3D_J~gg.
\ee
The decay widths of a virtual quark $Q^*$ to color-singlet quarkonium state
$^3D_J$ by gluon fragmentation can be evaluated via
\begin{eqnarray}
\Gamma(Q^*\rightarrow Q g^*; g^*\rightarrow ^3\!D_J~gg)=\int
\limits^s_{\mu^2_{min}}\!d\mu^2~\Gamma(Q^*\rightarrow Q g^*(\mu))\cdot 
P(g^*\rightarrow ^3\!D_J~gg),
\end{eqnarray}
where $s$ is the invariant mass squared of $Q^*$; $\mu$ is the virtuality 
of the gluon, and its minimum value squared $\mu_{min}^2=12 m_Q^2$ corresponding 
to the infrared cutoff as discussed below; 
$P$ is the decay distribution defined as
\begin{equation}
  P(g^*\rightarrow AX)\equiv\frac{1}{\pi\mu^3 }\Gamma(g^*\rightarrow AX).
\end{equation}
The calculations of $\Gamma(g^*\rightarrow AX)$ are lengthy but straightforward
 \cite{gf}.
So, the fragmentation functions can be calculated
\begin{eqnarray}
\label{4y}
D_{g^*\rightarrow^3D_J}(z,2m_Q,s)=\frac{d\Gamma(Q^*\rightarrow ^3\!D_J~gg~Q)/dz}
{\Gamma(Q^*\rightarrow Q g )},
\end{eqnarray}
where $z\equiv\frac{2P\cdot k}{\mu^2}=2-x_1-x_2$. At high energy limit, the  
interaction energy $s$ goes up to infinity, then the definition of $z$ here 
is identical with that in Ref.[2]  multiplied by a factor of two
and the fragmentation functions decouple from any specific gluon splitting
processes, which just reflects the universal spirit of fragmentation. 
The fragmentation function of
Eq.(\ref{4y}) is evaluated at the renormalization scale $2m_Q$, which
corresponds to the minimum value of the invariant mass of the 
virtual gluon. 

The calculation of color-octet fragmentation functions in
$g^*\rightarrow{}^3\!D_J({}^3S_1,\b 8)$ processes is
trivial. They may be obtained directly from color-octet
$g^*\rightarrow J/\psi({}^3\!S_1,\b 8)$ process \cite{surplus},
\begin{eqnarray}
\label{5x}
D_{g^*\rightarrow^3\!D_J}(z,2m_Q)=\frac{\pi\alpha_s(2m_Q)}{24 m_Q^2}\delta(1-z)
\< {\cal O}_8^{^3D_J}(^3S_1)\> .
\end{eqnarray}

For the numerical calculations, we choose
$$m_c=1.5~GeV,~ m_b=4.9~GeV,~ \alpha_s(2m_c)=0.26,~ \alpha_s(2m_b)=0.19,$$
\vskip -0.9cm
\begin{eqnarray}
\label{aaaa}
|R^{\prime\prime}_{(c\bar c)}(0)|^2=0.015~ GeV^7,~
|R^{\prime\prime}_{(b\bar b)}(0)|^2=0.637~ GeV^7.
\end{eqnarray}
And then, for color-singlet gluon fragmentation we obtain
\begin{eqnarray}
\label{7}
D^{(1)}_{g^*\rightarrow^3D_1(c\bar c)}=5.6\times 10^{-8},~
D^{(1)}_{g^*\rightarrow^3D_2(c\bar c)}=3.1\times 10^{-7},\nonumber\\
D^{(1)}_{g^*\rightarrow^3D_3(c\bar c)}=2.2\times 10^{-7},~
D^{(1)}_{g^*\rightarrow^3D_1(b\bar b)}=2.5\times 10^{-10},\nonumber\\
D^{(1)}_{g^*\rightarrow^3D_2(b\bar b)}=1.4\times 10^{-9},~
D^{(1)}_{g^*\rightarrow^3D_3(b\bar b)}=9.9\times 10^{-10}.
\end{eqnarray}

For color-octet gluon fragmentation, the fragmentation 
probabilities are proportional to the nonperturbative matrix elements 
$\< {\cal O}_8^{^3D_J}(^3S_1)\> $ which
have not been extracted out from experimental data, nor from the Lattice
QCD calculations. Based upon the NRQCD 
velocity scaling rules, here we tentatively assume \cite{gf} 
\begin{eqnarray}
\label{9}
\< {\cal O}_8^{^3D_2(c\bar c)}(^3S_1) \> \approx 
\< {\cal O}_8^{\psi'}(^3S_1)\> =4.6\times 10^{-3}~GeV^3
\end{eqnarray}
and further extend this relation to the $b\bar b$ system
\begin{eqnarray}
\label{10}
\< {\cal O}_8^{^3D_2(b\bar b)}(^3S_1)\> \approx 
\< {\cal O}_8^{\Upsilon'}(^3S_1)\> =4.1\times 10^{-3}~GeV^3.
\end{eqnarray}
Note that the NRQCD scaling rules tell us that the two matrix elements in (8)
(or (9)) are of the same order, but not necessarily equal, therefore the assumed
relations (\ref{9}) and (\ref{10}) certainly possess uncertainties
to some extent. However, from the calculated results below we are confident that 
it will not destroy our major conclusion.
From the approximate heavy quark spin symmetry relation, we have
\begin{eqnarray}
\label{11}
\< {\cal O}_8^{^3D_1}(^3S_1)\> \approx \frac{3}{5}
\< {\cal O}_8^{^3D_2}(^3S_1)\> \approx \frac{5}{7} 
\< {\cal O}_8^{^3D_3}(^3S_1)\>
\end{eqnarray}
for both $b\bar{b}$ and $c\bar{c}$ systems.

Using Eqs.(5), (\ref{aaaa}), and (\ref{9})-(\ref{11}), we readily have
\begin{eqnarray}
\label{12}
D^{(8)}_{g^*\rightarrow^3D_1(c\bar c)}=4.2\times 10^{-5},~
D^{(8)}_{g^*\rightarrow^3D_2(c\bar c)}=7.0\times 10^{-5},\nonumber\\
D^{(8)}_{g^*\rightarrow^3D_3(c\bar c)}=9.7\times 10^{-5},~
D^{(8)}_{g^*\rightarrow^3D_1(b\bar b)}=2.5\times 10^{-6},\nonumber\\
D^{(8)}_{g^*\rightarrow^3D_2(b\bar b)}=4.2\times 10^{-6},~
D^{(8)}_{g^*\rightarrow^3D_3(b\bar b)}=5.9\times 10^{-6}.
\end{eqnarray}
Comparing (\ref{12}) with (\ref{7}), we come to an 
anticipated conclusion that  at the Tevatron 
the gluon fragmentation probabilities through color-octet intermediates
to spin-triplet D-wave charmonium and bottomonium states are over $2\sim 4$ 
orders of 
magnitude larger than that of color-singlet processes. 
As a result, the production rates of ${}^3D_J$ charmonium states are about the same
amount as $\psi'$ production rates. Compared with the 
$\psi'$ production at the Tevatron, the gluon fragmentation color-octet
process plays an even more important role in the $^3D_J$ quarkonium production,
and it also 
gives production probabilities larger than the quark fragmentation 
process \cite{qf}.

\section{In $Z^0$ Decays}

For the $D$-wave heavy quarkonium production in $Z^0$ decays, the dominant process
is $Z^0\rightarrow {}^3D_J q\bar q$. This is similar to $Z^0\rightarrow \psi q\bar q$ discussed in Ref.[11].
Here $q$ represents $u,~d,~s,~c~{\rm or}~b$ quarks.
In the limit $m_q=0$, we readily have 
\begin{eqnarray}
\label{a10}
\nonumber
\Gamma(Z \rightarrow ^3\!\! D_{J} q\bar q ) & = &\Gamma(Z\rightarrow q\bar q)
\frac{\alpha_s^2(2m_c)}{36}\frac{\< O^{^3D_{J}}_8(^3S_1)\> }
{m_c^3}\big\{ 5(1-\xi^2)
-2\xi \ln \xi\\ 
\nonumber
&+& \big[2Li_2(\frac{\xi }{1+\xi })
 - 2 Li_2(\frac{1}{1+\xi })\\
&-& 2\ln(1+\xi )\ln\xi  + 3\ln\xi  + \ln^2 \xi \big]
(1+\xi)^2 \big\},
\end{eqnarray}
where $Li_2(x)=-\int\limits_0^x dt ~{\rm ln}(1-t)/t$ is the Spence function.
The calculation with physical masses, say, e.g. $m_b=5~GeV$, has also 
been performed by us, which does not show much difference from the case of
$m_q=0$.

From Eq.(\ref{a10}) (with slight modification due to nonzero $m_q$), 
we can get the branching ratios of 
$Z^0\rightarrow ^3\!D_J q\bar q$.
Summing over all the quark flavors ($q=u,~d,~s,~c,~b$) with their physical masses, 
we obtain the fraction ratios
\begin{eqnarray}
\frac{\Gamma(Z^0\rightarrow ^3\!\!D_1 q\bar q)}{\Gamma(Z^0\rightarrow q\bar q)}
=2.0\times 10^{-4},
\frac{\Gamma(Z^0\rightarrow ^3\!\!D_2 q\bar q)}{\Gamma(Z^0\rightarrow q\bar q)}
=3.4\times 10^{-4},
\frac{\Gamma(Z^0\rightarrow ^3\!\!D_3 q\bar q)}{\Gamma(Z^0\rightarrow q\bar q)}
=4.8\times 10^{-4}.
\end{eqnarray}

The color-singlet processes include the quark fragmentation contributions \cite{qf}
and the gluon fragmentation contributions.
For quark fragmentation, the branching
ratios of $^3D_J$ production 
in color-singlet processes are $2.3\times 10^{-6}$,
$3.6\times 10^{-6}$, $1.7\times 10^{-6}$ for $J=1,~2,~3$, respectively. 
The gluon fragmentation processes are more complicated.
For in the most important kinematic region 
the virtual gluon is nearly on its massshell,
$^3D_J$ production in the gluon fragmentation 
color-singlet process may be separated into $Z^0
\rightarrow q\bar q g^*$ and $g^*\rightarrow ^3\!\!D_J gg$.
The decay widths of $Z^0$ to color-singlet charmonium state $^3D_J$ 
by gluon fragmentation can be evaluated via
\begin{equation}
\Gamma(Z^0\rightarrow q\bar q g^*;g^*\rightarrow ^3\!D_J gg)=
  \int \limits_{\mu_{min}^2}^{M_Z^2} d\mu^2 \Gamma(Z^0\rightarrow q\bar q g^*)
    P(g^*\rightarrow^3\!D_J gg),
\end{equation}
where the cutoff $\Lambda=m_c$ is transformed into a lower limit on $\mu^2_{min}=
12 m_c^2$.
Summing over all the flavors $q~(q=u,d,s,c,b)$, we obtain
the fraction ratios for the production of ${}^3D_J$ quarkonium
are $4.3\times 10^{-7}$, $2.1\times 10^{-6}$, $1.2\times 10^{-6}$
for $J=1,~2,~3$, respectively.

\section{$D$-wave Charmonia Production in $B$ Decays}

We first calculate the color-singlet contribution to ${}^3D_1$ charmonium
production in $b$ decays.
The matrix elements $\< 0|(c\bar c)_{V-A}|{}^3D_1\> $
can be calculated as \cite{bdw},
\bea
\< 0|(c\bar c)_{V-A}|{}^3D_1\> &=&\epsilon_\mu\frac{20\sqrt{3}}{\sqrt{2\pi}}
        \frac{R_D^{\prime\prime}(0)}{M^3\sqrt{M}}.
\eea
So the partial width for color-singlet $\delta^c_1$ production in $b$ decays
\footnote{Here, we denote the physical $D$-wave charmonium state
as $\delta^c_J$, and ${}^3D_J$ represents the leading part of the Fock state compenents of
$\delta^c_J$.} will be
\bea
\Gamma(b\to c\bar c({}^3D_1^{(1)})+s,d\to\delta^c_1+X)
        &=&\frac{5}{9}\frac{\langle {\cal O}_8^{\delta^c_1}({}^3D_1)\> }{M_c^6}
        (2C_+-C_-)^2(1+\frac{8M_c^2}{M_b^2})\hat\Gamma_0.
\eea

The color-octet contributions to $\delta^c_J$ production are similar to that
to the $J/\psi$ production \cite{bsw}
\bea
\Gamma(b\to (c\bar c)_8+s,d\to\delta^c_J+X)&=&\big (
        \frac{\< {\cal O}_8^{\delta^c_J}({}^3S_1)\> }{2M_c^2}+
        \frac{\< {\cal O}_8^{\delta^c_J}({}^3P_1)\> }{M_c^4}\big )\\
        &&\times(C_++C_-)^2(1+\frac{8M_c^2}{M_b^2})\hat\Gamma^0.
\eea

The predicted branching fractions for $D$-wave charmonium states are
\bea
\label{d2}
B(B\to \delta^c_1+X)=0.28\%,~B(B\to \delta^c_2+X)=0.46\%,~
B(B\to \delta^c_3+X)=0.65\%.
\eea
We can easily find that the relative production rates predicted above are 
$\delta^c_1:\delta^c_2:\delta^c_3=3:5:7$. If we do
not take into account the color-octet mechanism the relative rates would be
$\delta^c_1:\delta^c_2:\delta^c_3=1:0:0$.

Comparing the predicted production rate of $2^{--}$ $D$-wave charmonium (\ref{d2})
with that of $\psi^\prime$, we can
see that the former has the same amount as the latter.
A similar discussion on the $D$-wave charmonium production in $B$ decays
was also presented in Ref.[12].

\section{Conclusions}

In conclusion, we have shown that in the production of $D$-wave heavy
quarkonia, the color-octet processes are always dominant. Their contributions
are orders of magnitude larger than the color-singlet contributions.
According to the NRQCD velocity scaling rules, the production
rate of $D$-wave quarkonium is the same order as that of $\psi^\prime$.
Among the three triplet states of $D$-wave charmonium, $\delta^c_2$ is the
most prominent candidate to be discovered firstly. It is a narrow resonance,
and the branching fraction of the decay mode $J/\psi \pi^+\pi^-$ is estimated 
to be \cite{z0}
\be
B(\delta^c_2\to J/\psi \pi^+\pi^-)\approx 0.12,
\ee
which is smaller than that of $B(\psi^{\prime}\to J/\psi \pi^+\pi^-)$ by only 
a factor of $3$.
Therefore, $2^{--}$ $D$-wave charmonium should be observable at the Tevatron via
color-octet gluon fragmentation, and at the CLEO via $B$ decays.
This would provide a crucial test for the color-octet production
mechanism.
We expect the experimental data will be soon obtained.

\section*{Acknowledgments}
This work was supported in part by the National Natural Science Foundation
of China, the State Education Commission of China, and the State Commission
of Science and Technology of China.
One of us (K.T.C.) would like to thank Prof.S.K. Kim-Ewha Womans University,
Prof.C. Lee-Seoul National University, and Prof. K.Kang-Brown University
for their outstanding organization and hospitality during this conference.

\end{document}